\documentclass[pop,fleqn]{w-art}
\usepackage{times}
\usepackage{w-thm}
\usepackage{amsmath}
\usepackage{amsfonts} 
\usepackage{amssymb} 

\theoremstyle{plain}

\theoremstyle{definition}

\usepackage{graphicx}
\def\ee{\mathrm{e}}

\newcommand{\Tr}{\mbox{Tr}}
\newcommand{\tr}{\mathrm{tr}}

\def\one{{\rm 1\kern -.9mm l}}

\def\ii{{\mbox{i}}}

\newcommand{\uno}{\mbox{1\!\negmedspace1}}

\newcommand{\smod}{S_{\rm mod}({q},{\tilde q};{\cal M}_{k})}
\newcommand{\smodh}{S_{\rm mod}({q},{\tilde q};{\widehat{\cal M}_{k}})}

\newcommand{\Lhol}{\Lambda_{\scriptscriptstyle\mathrm{hol}}}
\newcommand{\Lholt}{\widehat{\Lambda}_{\scriptscriptstyle\mathrm{hol}}}
\newcommand{\Gammab}{\boldsymbol{\Gamma}}
\begin{document}
\DOIsuffix{theDOIsuffix}
\Volume{??}
\Issue{?}
\Month{??}
\Year{????}
\pagespan{1}{}



\title[(D)-instantons in brane worlds]{(D)-instanton effects in magnetized
brane worlds
}


\author[M. Bill\'o]{Marco Bill\'o\inst{1}%
  \footnote{Corresponding author\quad E-mail:~\textsf{billo@to.infn.it},
            Phone: +0039\,011\,670\,7213,
            Fax: +0039\,011\,670\,7213}}
\address[\inst{1}]{Dip. di Fisica Teorica, Universit\`a di Torino
and I.N.FN., sez. di Torino\\ 
Via Pietro Giuria 1, 10125 Torino (ITALY)}
\begin{abstract}
We consider systems of magnetized D9
branes on orbifolds supporting $\mathcal{N}=1$ gauge
theories. In such realizations, the matter multiplets arise from ``twisted''
strings connecting different stacks of branes. The introduction of Euclidean
5-branes wrapped on the six-dimensional compact space  leads to
instanton effects. 
We examine the interplay between the annuli diagrams
with an E5 boundary and the holomorphicity properties  of the effective
low-energy supergravity action which describes the system, including its
instanton corrections. This talk is mostly based on \cite{Billo:2007py}.
\end{abstract}
\maketitle                   






\section{Introduction}
A popular and promising scenario for embedding phenomenologically appealing gauge
and matter content in String Theory entails considering systems of magnetized
D9 branes in type IIB Superstring, compactified on suitable
six-dimensional manifolds; the T-dual description in terms of intersecting D6 branes in type IIA
can also be used \cite{Blumenhagen:2006ci}.
Such constructions support, in the uncompactified
$\mathbf{R}^{1,3}$ space, supersymmetric gauge theories with chiral matter
and interesting phenomenology. In particular, replica families arise naturally
from multiple intersections, and different coupling constants can be tuned for
the various gauge group factors. The low energy theory describing such
four-dimensional theories is provided by  supergravity coupled to vector and
matter multiplets, and it can be derived directly from string
amplitudes. Having and explicit string description makes it possible to 
investigate whether novel stringy effects, both perturbative and
non-perturbative, might contribute to the effective action.

In this respect, it is by now clear that Euclidean E5 branes
wrapped on the same cycle occupied by one of the stacks of D9 branes, which are
point-like in $\mathbf{R}^{1,3}$, correspond to instantonic configurations of
the gauge theory living on the D9 stack. 

This is entirely analogous to the well-known D3/D(-1) system:
the ADHM construction is naturally reproduced by the open strings attached to
the instantonic branes \cite{Witten:1995im,Douglas:1995bn},
while the non-trivial instanton profile of the gauge field is produced by the
emission of gauge field vertices from mixed disks \cite{Billo:2002hm}.

E5 branes wrapped differently from the D9 branes are still
point-like in $\mathbf{R}^{1,3}$ but do not correspond to ordinary instantons
configurations \cite{Beasley:2005iu}. Still they can, in certain cases, give important non-perturbative,
stringy contributions to the effective action, .e.g.,  Majorana masses for
neutrinos \cite{Blumenhagen:2006xt,Ibanez:2006da}, moduli stabilizing terms and others, 
with a growing literature which impossible to quote here extensively.
These ``exotic'' instantonic branes are thus potentially crucial for string based phenomenology.

Our aim here is to clarify some aspects of the ``stringy instanton
calculus'', i.e., of the computation of Euclidean branes contributions.
We focus on ordinary instantons, but our considerations should be useful
for exotic instantons as well. We choose a toroidal compactification
 where string theory is calculable  and realize (locally) $\mathcal{N}=1$ gauge
SQCD on a system of D9-branes. We discuss then the contribution of 
E5 branes to the superpotential, analyzing the r\^ole of annuli
bounded by E5 and D9 branes in giving these terms  
suitable holomorphicity properties.

\section{The set-up}
We take as internal space the orbifold 
\begin{equation}
\frac{\mathcal{T}_2^{(1)}\times\mathcal{T}_2^{(2)}
\times\mathcal{T}_2^{(3)}}{\mathbb{Z}_2\times \mathbb{Z}_2}~,
\end{equation}
whose K\"ahler  and complex structures, $T^{(i)}$ and $U^{(i)}$
respectively with $i=1,2,3$ referring to the three tori, parametrize the string
frame metric and the $B$ field. We decompose the 10-d string fields into 4d and
6d components as $X^M\to (X^\mu,Z^i)$ and $\psi^M\to (\psi^\mu,\Psi^i)$, with
$
Z^i = \textstyle{\sqrt{\frac{T_2^{(i)}}{2U_2^{(i)}}}} (X^{2i+2} + U^{(i)}
X^{2i+3})$; the spin field decomposition we write as $S^{\dot{\mathcal{A}}}
\to (S_\alpha S_{---}, S_\alpha S_{-++},\ldots,S^{\dot\alpha} S^{+++},\ldots)$.
The three non-trivial orbifold elements $h_i$ act as follow: 
\begin{equation}
\label{frac2}
\begin{aligned}
h_1:~(Z^1,Z^2,Z^3) &\rightarrow (Z^1,-Z^2,-Z^3)~,\\
h_2:~(Z^1,Z^2,Z^3) &\rightarrow (-Z^1,Z^2,-Z^3)~,\\
h_3:~(Z^1,Z^2,Z^3) &\rightarrow (-Z^1,-Z^2,Z^3)~.
\end{aligned}
\end{equation}
The $\mathcal{N}=1$ supergravity describing the low energy theory of the closed
string fields in this background can be expressed in terms of the fields
$s,t^{(i)},u^{(i)}$, with \cite{Blumenhagen:2006ci}
\begin{equation}
\begin{aligned}
&\mathrm{Im}(s) \equiv s_2 = \frac{1}{4\pi}\,\ee^{-\phi_{10}}\,
T_2^{(1)}T_2^{(2)}T_2^{(3)}~,
\\
&\mathrm{Im}(t^{(i)}) \equiv t_2^{(i)} = \ee^{-\phi_{10}} T_2^{(i)}~,
~~~
u^{(i)} = u_1^{(i)} + \ii\, u_2^{(i)} = U^{(i)}~,
\end{aligned}
\label{stu}
\end{equation}
whose bulk K\"ahler potential reads \cite{Antoniadis:1996vw}
\begin{equation}
K = -\log (s_2) -\sum_{i=1}\log(t_2^{(i)}) -
\sum_{i=1} \log(u_2^{(i)})~.
\end{equation}

We place now in this background a stack of $N_a$ fractional D9 branes (which we
will call ``color branes'' and denote as {9$a$} branes).
The massless spectrum of 9$a$/9$a$ strings gives rise,  in
$\mathbf{R}^{1,3}$,  to the  $\mathcal{N}=1$ vector multiplet for the
gauge group $\mathrm{U}(N_a)$. The gauge coupling is given at tree level by%
\footnote
{It is however the Wilsonian coupling $1/{\tilde g}_a^2$ which has to be the imaginary part of a chiral multiplet, and it can in principle be corrected by sigma-model anomalies
\cite{Derendinger:1991hq}, so that
\begin{equation}
\label{wcorr}
 \frac{1}{{\tilde g}_a^2} =  s_2 = \frac{1}{{g}_a^2} - \frac{\delta}{8\pi^2}~. 
\end{equation}
Only a shift $\delta^{(0)}$ could however play a role in our later discussion, see \cite{Billo:2007py} for further details.
}
\begin{equation}
\frac{1}{g_a^2} =
\frac{1}{4\pi}\,\ee^{-\phi_{10}}\,T_2^{(1)}T_2^{(2)}T_2^{(3)}
= s_2~.
\end{equation}

We add a further stack of D9-branes (``flavor branes'' {9$b$}) 
with quantized magnetic fluxes
${f_b^{(i)}} = m_b^{(i)}/n_b^{(i)}$
and in a different orbifold representation.
(Bulk) supersymmetry requires $\nu_b^{(1)}-\nu_b^{(2)}-\nu_b^{(3)}=0$, where 
\begin{equation}
{f_b^{(i)}}/T_2^{(i)} = \tan \pi{\nu_b^{(i)}}~,~~~~~~~\mbox{with}~~~~
0\leq \nu_b^{(i)}< 1~.
\label{nui}
\end{equation}
Open strings stretched between the two stacks of branes
(9$a$/9$b$ strings) are twisted by the relative 
angles 
\begin{equation}
\nu^{(i)}_{ba}=\nu^{(i)}_{b} -\nu^{(i)}_{a}~.
\end{equation}
If  $\nu_{ba}^{(1)}-\nu_{ba}^{(2)}-\nu_{ba}^{(3)}=0$, this sector
is supersymmetric: massless modes fill up a chiral multiplet $q_{ba}$ in
the anti-fundamental representation $\bar N_a$ of the color group.
The degeneracy of this chiral multiplet is $N_b |I_{ab}|$,
where {$I_{ab}$} is the number of Landau levels for the $(a,b)$ ``intersection''
\begin{equation}
I_{ab} = \prod_{i=1}\big(m_a^{(i)}n_b^{(i)}-
m_b^{(i)}n_a^{(i)}\big)~.
\label{iab}
\end{equation}
We can now introduce a third stack of 9$c$ branes such that we get a chiral
multiplet $q_{ac}$ in the fundamental representation $N_a$ and that 
\begin{equation}
N_b |I_{ab}|=N_c |I_{ac}| \equiv N_F~.
\end{equation}
This gives a (local) realization of $\mathcal{N}=1$ SQCD, since we have the same
number $N_F$ of fundamental and anti-fundamental chiral multiplets, respectively
denoted by {$q_f$} and {${\tilde q}^f$.
The kinetic terms of the scalars $q_f$ sitting in the chiral multiplets, as 
evaluated from disk amplitudes, take a canonical form. 
This is not the case in the 
supergravity Lagrangian, where fields $Q_f$ with different normalizations
are employed whose kinetic terms take the form
\begin{equation}
\sum_{f=1}^{N_F}
\Big\{ {K_Q}\,D_\mu {Q^{\dagger f}} \,D^\mu {{Q}_f}
 + {K_{\tilde Q}}\, D_\mu {{\tilde Q}^f}\,D^\mu {{\tilde Q}^\dagger_f}
\Big\}~.
\label{kinQ}
\end{equation}
The ``string'' and ``supergravity'' fields are thus related via the K\"ahler
metrics:
\begin{equation}
{q = \sqrt{K_{Q}^{}}\,Q}\, ,~~~
{\tilde q = \sqrt{K_{\tilde Q}}\,\tilde Q}~.
\label{qQ}
\end{equation}

\subsection{Non-perturbative sectors from $E5$ branes}
In the string realization of SQCD described above, we can introduce
instantonic sectors of the $\mathrm{U}(N_a)$ gauge theory by adding stacks of
$k$ E5 branes whose internal part coincides with the D9$a$. Notice that these
branes would represent exotic instantons for the gauge theories on the D9
branes of type $b$ and $c$. 
We have new types of open strings: E$5_a$/E$5_a$ (neutral sector), 
D$9_a$/E$5_a$ 
(charged sector),
D$9_b$/E$5_a$ or E$5_a$/D$9_c$
(flavored sectors, twisted).
The states of these strings carry no momentum in space-time. They
represent moduli, not fields. We shall indicate them collectively
by $\mathcal{M}_k$; their spectrum and properties are summarized  in table
\ref{table:mod}. Notice that charged or neutral moduli can have
Kaluza-Klein momentum in (some of) the internal tori.

\begin{table}
\begin{tabular}{cc|cccc}
\multicolumn{2}{c}{Sector}  & ADHM  & Meaning & Chan-Paton & Dimension\\
\hline
$\phantom{\vdots}{5_a/5_a}$ & NS & $a_\mu$ & centers & adj.
$\mbox{U}(k)$ &
(length)\\
 & &  $D_c$ & Lagrange mult. & $\vdots$ & (length)$^{-2}$\\
 &  R & ${M}^{\alpha}$ &  partners &  $\vdots$ & (length)$^{\frac12}$\\
 &    & $\lambda_{\dot\alpha}$ & Lagrange mult.  & $\vdots$ &
(length)$^{-\frac32}$ \\
\hline
$\phantom{\vdots}{9_a}/{5_a}$ & NS &  ${w}_{\dot\alpha}$ & sizes
& $N_a \times
\overline{k}$
& (length)\\
${5_a}/{9_a}$ &  & ${\bar w}_{\dot\alpha}$ & $\vdots$ & $k\times
\overline{N}_a$ &
$\vdots$\\
${9_a}/{5_a}$ & R & ${\mu}$ & partners & $N_a \times
\overline{k}$ & (length)$^{\frac12}$\\
${5_a}/{9_a}$ &  & ${\bar \mu}$ & $\vdots$ &$k\times
\overline{N}_a$ &  $\vdots$\\
\hline
$\phantom{\vdots}{9_b}/{5_a}$ & R & ${\mu}^\prime$ & flavored &
${N}_F\times
\overline{k}$
&  (length)$^{\frac12}$\\
${5_a}/{9_c}$ &  & ${\tilde \mu}^\prime$ & $\vdots$ & ${k}
\times\overline{N}_F
$ & $\vdots$\\
\end{tabular}
\caption{The spectrum of moduli, labeled in accordance with their r\^ole in the
ADHM construction.}
\label{table:mod}
\end{table}

Among the neutral moduli we have  the center of mass position
$x_0^\mu$ and its fermionic  partner $\theta^\alpha$,
related to the supersymmetries broken by the E5$a$ branes, which correspond to
the singlet part:
 \begin{equation}
{a}^\mu = x_0^\mu\,\uno_{k\times k} + y^\mu_c\,T^c\quad,\quad
{M}^{\alpha}=\theta^{\alpha}\,\uno_{k \times k } +
{\zeta}^{\alpha}_c\,T^c~.
\label{xtheta}
\end{equation}

In the flavored sectors one has fermionic  zero-modes only:
the  $\mu'_f$ from the D$9_b$/E$5_a$ sector and the ${\tilde \mu}'{}^f$
from the E$5_a$/D$9_c$ sector.

\section{The stringy instanton calculus}
In presence of Euclidean branes, the dominant 
contributions to correlators of gauge/matter fields arise from
products of one-point functions, connected by the integration over the moduli. 
The effective action for the gauge/matter fields is thus
obtained by the “functional” integral over the instanton moduli of the exponential of all
diagrams with at least part of their boundary on the E5a branes, possibly with insertions of
moduli and gauge/matter fields\cite{Polchinski:1994fq}-\cite{Green:1998yf},\cite{Billo:2002hm,Blumenhagen:2006xt}. 
In the semi-classical approximation, only disk diagrams and annuli (the latter with no insertions) are retained.

In the case at hand, taking into account the dependence from
the scalars $q$, $\tilde q$, the instantonic disk diagrams account for the shifted moduli action
\begin{equation}
\label{smodex}
\begin{aligned}
S_{\rm mod}(q,\tilde q;\mathcal{M}_k)& = {\rm tr}_k  \Big\{
\ii D_c\Big({\bar
w}_{\dot\alpha}(\tau^c)^{\dot\alpha}_{\,\dot\beta}w^{\dot\beta}
+ \ii \bar\eta_{\mu\nu}^c \big[{a}^\mu,{a}^\nu\big]\Big)
\\& - \ii
{\lambda}^{\dot\alpha}\Big(\bar{\mu}{w}_{\dot\alpha}+
\bar{w}_{\dot\alpha}{\mu}  +
\big[a_\mu,{M}^{\alpha}\big]\sigma^\mu_{\alpha\dot\alpha}\Big)\Big\}
\\
& + \tr_k\sum_{f} \Big\{ {\bar w}_{\dot\alpha}
\big[{q^{\dagger f}{q}_f}
 + {\tilde q^f \tilde{q}^\dagger_f}\big]
 w^{\dot\alpha} - \frac{\ii}{2}\, {\bar \mu}\,
{q^{\dagger f}} \mu'_f +  \frac{\ii}{2}\,
{{\tilde\mu}'}{}^f{\tilde{q}^\dagger_f}\,
\mu\Big\}~.
\end{aligned}
\end{equation}
There are other relevant diagrams which involve the superpartners of ${q}$ and
${\tilde q}$ and are related to the above by susy Ward identities. The
complete result is obtained by letting
\begin{equation*}
 {q}(x_0)\,,~{\tilde q}(x_0) ~ \to ~  {q}(x_0,\theta) \,,~{\tilde q}(x_0,\theta)
\end{equation*}
in $\smod$. Using this ingredient, the effective action in the Higgs branch
takes the form
\begin{equation}
S_{k}=  {\cal C}_k ~\ee^{-\frac{8 \pi^2}{g_a^2}\,k}
\ee^{\mathcal{A}^\prime_{5_a}} \int d{\mathcal M}_{k}~
\ee^{-\smod}~.
\end{equation}
In $\mathcal{A}^\prime_{5_a}$ the contribution of zero-modes running in the
loop is suppressed because they are already explicitly integrated over
\cite{Blumenhagen:2006xt,Abel:2006yk,Akerblom:2006hx,Billo:2007sw}.
${\cal C}_k$ is a normalization factor,  determined up to numerical
constants by counting the dimensions, measured in units of $\alpha'$,
of the moduli ${\mathcal M}_{k}$:
\begin{equation}
{\mathcal C}_k = \big({\sqrt {\alpha'}}\big)^{-(3N_a-N_F)k}\, (g_a)^{-2N_ak}~.
\label{ck}
\end{equation}
Notice the appearance of the $\beta$-function coefficient. $b_1= N_a-N_F$. 

In $\smod$, the superspace coordinates {$x_0^{\mu}$} and
{$\theta^{\alpha}$} appear only within the superfields
${q}(x_0,\theta)$, ${\tilde q}(x_0,\theta)$, $\ldots$. 
We can thus separate {$x,\theta$} from the other moduli $\widehat{\mathcal
M}_{k}$ writing
\begin{equation}
S_{k}= \int {d^4 x_0\, d^2\theta} ~W_{k}(q,\tilde q)
\label{Wk}
\end{equation}
in terms of the effective superpotential
\begin{equation}
W_{k}(q,\tilde q)=
{\cal C}_k ~\ee^{-\frac{8 \pi^2}{g_a^2}\,k}
\ee^{\mathcal{A}^\prime_{5_a}}
\int d{\widehat{\mathcal M}_{k}}~
\ee^{-\smodh}~.
\end{equation}
This expression seems not to be holomorphic, as a superpotential should be.
Indeed, $\smodh$ explicitly depends  on {$q^\dagger$} and {${\tilde
q}^\dagger$}. This dependence disappears upon integrating over
$\widehat{\mathcal M}_{k}$ as a consequence of the cohomology properties of the
integration measure on the instanton moduli space 
\cite{Hollowood:2002ds,Dorey:2002ik,Billo:2006jm}.
However, we have to re-express the result in terms of the supergravity fields 
{$Q$} and {${\tilde Q}$} via the rescalings in eq. (\ref{qQ}) and this
reintroduces non-holomorphicity. 
At the same time, we have to use the holomorphic dynamically generated scale $\Lhol$, obtained by
integrating the Wilsonian $\beta$-function of the $\mathcal{N}=1$ SQCD \cite{Novikov:1983uc,Dorey:2002ik}:
\begin{equation}
\Lambda_{\scriptscriptstyle\mathrm{hol}}^{b_1}=
(\sqrt{\alpha'})^{-b_1} \ee^{-\frac{8\pi^2}{{\tilde g}_a^2}}~,
\label{holPV}
\end{equation}
where the holomorphic coupling $\tilde g_a$ of (\ref{wcorr}) appears.
Finally,  $\mathcal{A}^\prime_{5_a}$
can introduce a non-holomorphic dependence on the complex and K\"ahler
structure moduli of the compactification space.
Our aim is to consider the interplay of all these observations; for this we
need the explicit expression of the mixed annuli term
$\mathcal{A}^\prime_{5_a}$, which we will discuss shortly. For the time being,
let us consider the one-instanton sector, $k=1$. In this case,
the integral over the moduli can be carried out explicitly.
Balancing the fermionic zero-modes requires that $N_F = N_a - 1$ and the end
result is \cite{Dorey:2002ik} (see also Refs.
\cite{Akerblom:2006hx,Argurio:2007vq}).
\begin{equation}
\label{ADS}
W_{k=1}(q,\tilde q)
= {\cal C}_k ~\ee^{-\frac{8 \pi^2}{g_a^2}\,k}
\ee^{\mathcal{A}^\prime_{5_a}}\,
\frac1{\det\big({\tilde q} {q}\big)}~.
\end{equation}
This superpotential has basically the same form as the ADS/TVY superpotential
\cite{Taylor:1982bp,Affleck:1983mk}
except for the prefactors discussed above:  we shall see how these factors
precisely conspire to give an holomorphic expression in the supergravity variables
{$Q$} and {${\tilde Q}$}.

\section{Instanton annuli and threshold corrections}
The amplitude $\mathcal{A}_{5_a}$ is a sum of cylinder amplitudes with 
a boundary on the {E5$a$} (both orientations)
\begin{figure}[htb]
 \begin{center}
\begin{picture}(0,0)%
\includegraphics{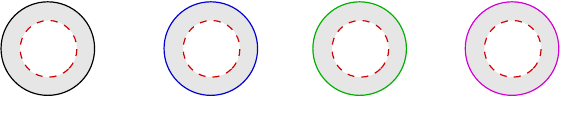}%
\end{picture}%
\setlength{\unitlength}{1492sp}%
\begin{picture}(7113,1621)(612,-1201)
\put(2046,-291){\makebox(0,0)[lb]{\smash{{$=$}}}}
\put(4086,-291){\makebox(0,0)[lb]{\smash{{$+$}}}}
\put(5991,-291){\makebox(0,0)[lb]{\smash{{$+$}}}}
\put(1081,-1186){\makebox(0,0)[lb]{\smash{{$\mathcal{A}_{5_a}$}}}}
\put(3006,-1186){\makebox(0,0)[lb]{\smash{{$\mathcal{A}_{5_a;9_a}$}}}}
\put(4851,-1186){\makebox(0,0)[lb]{\smash{{$\mathcal{A}_{5_a;9_b}$}}}}
\put(6831,-1186){\makebox(0,0)[lb]{\smash{{$\mathcal{A}_{5_a;9_c}$}}}}
\end{picture}%
\end{center}
\caption{Mixed annuli diagrams appearing in the stringy instanton calculus}
\label{fig:annuli}
\end{figure}
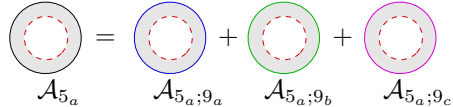
These amplitudes are both UV and IR divergent. The UV divergences (IR in the
closed string channel) cancel if tadpole cancellation assumed. The IR
divergence is regulated with a scale $\mu$.
As noticed in \cite{Abel:2006yk,Akerblom:2006hx}, there is a relation
between these instantonic annuli and the running
gauge coupling constant:
\begin{equation}
\label{a5tog}
\mathcal{A}_{5_a} = -\frac{8\pi^2 k}{g_a^2(\mu)}\,\Bigg|_{\mathrm{1-loop}}~.
\end{equation}
Indeed, in supersymmetric theories, mixed annuli compute the running coupling by
expanding around the instanton background, as discussed in \cite{Billo:2007sw}.

This relation is confirmed by the explicit computation of the annuli. Imposing
the appropriate boundary conditions and the GSO projection one starts from
\begin{equation}
\label{ma0}
\int _0^\infty \frac{d\tau}{2\tau}\left[
\Tr_{\mathrm{NS}}\left(P_{\mathrm{GSO}} \,P_{\mathrm{orb.}}\,
q^{L_0}\right)
- \Tr_{\mathrm{R}}\left(P_{\mathrm{GSO}}\,P_{\mathrm{orb.}}\,
q^{L_0}\right)\right]~.
\end{equation}

For $\mathcal{A}_{{5_a};{9_a}}$, KK copies of zero-modes
on internal tori $\mathcal{T}_2^{(i)}$ give a non-holomorphic dependence on
the K\"ahler and complex moduli (threshold correction) and one finds
\cite{Lust:2003ky,Akerblom:2007np} 
\begin{equation}
 \label{A5finaa}
 \mathcal{A}_{{5_a};{9_a}}  = - 8 \pi^2 k
 \Bigl[\frac{3N_a}{16\pi^2} \log(\alpha^\prime \mu^2) 
 + \frac{N_a}{16\pi^2}
 \sum_{i}
 \log \Big(U_2^{(i)}T_2^{(i)} (\eta(U^{(i)})^4\Big)\Bigr]~,
 \end{equation}

For $\mathcal{A}_{{5_a};{9_b}}$ and
$\mathcal{A}^\prime_{{5_a};{9_c}}$, the modes are twisted 
 and the result depends on the angles {$\nu_{ba}^{(i)}$} and
{$\nu_{ac}^{(i)}$}: 
\begin{equation}
\mathcal{A}_{{5_a};{9_b}} +
\mathcal{A}_{{5_a};{9_c}}  = 8\pi^2 k
\Bigl(\frac{N_F}{16\pi^2} \log(\alpha^\prime
\mu^2)  + \frac{N_F}{32\pi^2} \log
\left( \Gammab_{ba} \, \Gammab_{ac} \right) \Bigr)~,
\end{equation}
where 
\begin{equation}
\Gammab_{ba} = 
\frac{\Gamma(1-\nu_{ba}^{(1)})}{\Gamma(\nu_{ba}^{(1)})}
\frac{\Gamma(\nu_{ba}^{(2)})}{\Gamma(1 - \nu_{ba}^{(2)})}
\frac{\Gamma(\nu_{ba}^{(3)})}{ \Gamma(1 - \nu_{ba}^{(3)})}
\end{equation}
and $\Gammab_{ac}$ has an analogous expression.

Summing all types of annuli, we notice that the coefficient of the IR divergent
term $\log\alpha'\mu^2$ is proportional to the
{$\beta$-function coefficient} of SQCD {$b_1 = 3 N_a - N_F$}.

\section{Holomorphicity properties}
As stressed before, the instanton-induced correlators involve the primed part
$\mathcal{A}_{5a}^\prime$ 
of the mixed annuli, deprived of the the contributions of the zero-modes
running in the loop, which are responsible
for the IR divergences.
To isolate such contributions, we use the natural UV cut-off
of the low-energy theory, the Planck mass
\begin{equation}
\label{mp}
M_P^2\,=\,\frac{1}{\alpha'}\,{\rm e}^{-\phi_{10}}\,s_2~.
\end{equation}
We write then 
\begin{equation}
 \mathcal{A}_{5a} = -k \frac{b_1}{2} \log\frac{\mu^2}{M_P^2} + 
\mathcal{A}_{5a}^\prime~.
\end{equation}
With some algebra, and recalling the definition of the supergravity variables, we find
\begin{equation}
\label{a5p}
\mathcal{A}_{5_a}^\prime  =
- N_a \sum_{i=1}^3 \log \left(\eta(u^{(i)})^2 \right) + N_a \log g_a^2
+ \frac{N_a - N_F}{2} K 
+ \frac{N_F}{2} \log({\mathcal{Z}_{ba}} {\mathcal{Z}_{ac}})
\end{equation}
with (similarly for ${\mathcal{Z}_{ac}}$)
\begin{equation}
{\mathcal{Z}_{ba}} =  \big(4\pi\,s_{2}\big)^{-\frac14}\,
\big( t_{2}^{(1)} t_{2}^{(2)}  t_{2}^{(3)} \big)^{-\frac14}\,
\big( u_{2}^{(1)}   u_{2}^{(2)} u_{2}^{(3)}\big)^{-\frac12}
\,\big({\Gammab_{ba}}\big)^{\frac12}~.
\label{KQ}
\end{equation}

Let us focus now on the one instanton case $k=1$. When $N_F = N_a - 1$ we
found the effective superpotential of eq. (\ref{ADS}). 
Inserting the explicit form of the annuli, eq. (\ref{a5p}), and rewriting the resulting
expression in terms of the holomorphic scale $\Lhol$ of eq. (\ref{holPV}), we get
\begin{equation}
W_{k=1}({q},{\tilde q}) = 
\ee^{{K}/2}\,\prod_{i=1}^3\left(\eta(u^{(i)})^{-2 N_a}\right)\,
\Lhol^{2N_a+1}
\left({K_{ba}} {K_{ac}}\right)^{\frac{N_a-1}2}\,
\frac{1}{\det({\tilde q} {q})}~,
\end{equation}
where, taking into account a possible shift $\delta^{(0)}$ in the definition of the Wilsonian coupling, see eq. (\ref{wcorr}), we have
introduced   
\begin{equation}
\label{chicorr}
 K_{ab} = \chi_{ab} \mathcal{Z}_{ab}~,
\end{equation}
with the multiplicative corrections $\chi_{ab}$ satisfying 
\begin{equation}
 \label{restrchi}
\delta^{(0)} + \frac{N_F}{2}\log\chi_{b_a}\chi_{ac} = 0~.
\end{equation}

We can then make an holomorphic redefinition of the scale $\Lhol$ into
$\Lholt$ and rescale the chiral multiplet to their supergravity counterparts.
If we assume that the K\"ahler metrics for the chiral multiplets
are given by
\begin{equation}
{K_Q} = {K_{ba}}~,~~~
{K_{\tilde Q}} = {K_{ac}}
\end{equation}
we finally obtain an expression which fits perfectly in the low energy 
Lagrangian:
\begin{equation}
W_{k=1}({Q},{\tilde Q})
= 
\ee^{{K}/2}\,\
\Lholt^{2N_a+1} \frac{1}{\det({\tilde Q} {Q})}
\end{equation}
which, a part from the prefactor $\ee^{\frac K2}$, is  
holomorphic in the variables of the Wilsonian scheme.

The holomorphicity properties of the instanton-induced superpotential
suggest that the K\"ahler metric of chiral multiplets {$Q$} arising from twisted
{D9$_a$}/{D9$_b$} strings is given, a part from the multiplicative corrections $\chi_{ba}$ of eq. (\ref{chicorr}), by
\begin{equation}
{K_{Q}} =  \big(4\pi\,s_{2}\big)^{-\frac14}\,
\big( t_{2}^{(1)} t_{2}^{(2)}  t_{2}^{(3)} \big)^{-\frac14}\,
\big( u_{2}^{(1)}   u_{2}^{(2)} u_{2}^{(3)}\big)^{-\frac12}
\,\big({\Gammab_{ba}}\big)^{\frac12}~, 
\end{equation}
with 
\begin{equation}
{\Gammab_{ba}} = 
\frac{\Gamma(1-\nu_{ba}^{(1)})}{\Gamma(\nu_{ba}^{(1)})}
\frac{\Gamma(\nu_{ba}^{(2)})}{\Gamma(1 - \nu_{ba}^{(2)})}
\frac{\Gamma(\nu_{ba}^{(3)})}{ \Gamma(1 - \nu_{ba}^{(3)})}~.
\end{equation}
This is very interesting because
for {twisted} fields, the K\"ahler metric cannot be derived from
compactification of the DBI action. The part depending on the twists, namely
{$\Gamma_{ba}$}, is 
reproduced by a direct string computation~\cite{Lust:2004cx,Bertolini:2005qh}.
The prefactors, depending on the geometric moduli, are
more difficult to get directly: the present suggestion is welcome.

This expression of the K\"ahler metric can be checked against the known results
for Yukawa couplings of magnetized branes~\cite{Cremades:2003qj},
finding perfect consistency.

\subsection{Relation with the perturbative approach and the DKL formula}
We have discussed the relation between the instantonic annuli and the
running gauge coupling. In turn, there is a general relation of the 1-loop
corrections \cite{Dixon:1990pc,Kaplunovsky:1994fg} to the Wilsonian gauge coupling $f$:

\begin{equation}
\frac{1}{g^2(\mu)}
=\frac{1}{{\tilde g}^2}\frac{1}{8\pi^2}\left[\frac{b}2\,\log \frac{\mu^2}{M_P^2} - f^{(1)} - \frac{c}{2} K
+ T(G)\log \frac{1}{{\tilde g}^2} - \sum_r n_r T(r) \log K_r \right]~,
\label{kl}
\end{equation}
where $T_A$ are the generators of the gauge group, $n_r$ is the number of chiral multiplets in
the representation $r$ and 
\begin{equation}
\begin{aligned}
&T(r)\,\delta_{AB}=\Tr_r\big(T_AT_B\big)\quad,\quad
T(G)=T(\mathrm{adj})
\\
&b=3\,T(G)-\sum_r n_r\, T(r)\quad,\quad c=T(G)-\sum_r n_r\,T(r)~.
\end{aligned}
\end{equation}
This gives an interpretation for the {non-holomorphic terms} appearing in the
running coupling 
based on perturbative considerations. 

In the case of SQCD, one has $N_F$ chiral multiplets in the representations $N_a$ and
$\bar N_a$. Matching the DKL formula with the 1-loop result for
$1/g_A^2(\mu)$ one can identify the K\"ahler metrics {$K_Q$} and {$K_{\tilde
Q}$} of the chiral multiplets.
This determination, based on the holomorphicity of perturbative
contributions  to the effective action, is in full agreement with the expression
given before, derived from the holomorphicity of instanton
contributions.

\section{Remarks and conclusions}

The type of analysis reviewed here can be applied also in $\mathcal{N}=2$
toroidal models, where one can show that the instanton-induced superpotential is
in fact  {holomorphic} in the appropriate supergravity variables if one includes the
mixed annuli in the stringy instanton calculus \cite{Akerblom:2006hx,Billo:2007sw}.
In the present analysis, we have focused on the fact that, with respect to the
``color'' {D9$_a$} branes, the {E5$_a$} branes represent
ordinary instantons. For the gauge theories on the {D9$_b$} or the 
{D9$_c$}, they would be {exotic} instantons, whose status is not yet completely 
clear from the field theory viewpoint. The study of the {mixed annuli} and their relation to
holomorphicity can thus be  relevant for {exotic}, new {stringy effects} as
well.

\end{document}